# Research Approaches on Energy-aware Cognitive Radio Networks and Cloud based Infrastructures

Nikolaos Atzarakis
Department of Informatics Engineering
Technological Educational Institute of Crete
Heraklion, Crete, Greece
atza_nick@yahoo.com

*Abstract*—As years passes wireless networks were rapidly improved, introducing new applications and services, as well as important challenges for mobility support. This research field is new with many researchers and scientists making their proposals to optimize the provision of multiple services to the mobile users. In this context, this survey paper studies research approaches from the following topics: Cognitive Radio Networks, Interactive Broadcasting, Energy Efficient Networks, Cloud Computing and Resource Management.

## I. INTRODUCTION

The raise of mobility has make more needs for wireless connectivity and bandwidth. Users of wireless networks increased as years pass and make the need for more applications with less bandwidth and energy demands. In this paper we study and present related work categorized by topic.

- *Section II: Cognitive Radio Networks·* provides communication that exploits efficiently radio spectrum resources.
- *Section III: Interactive Broadcasting·* proposes a network architecture that exploits the terrestrial digital video broadcasting technology (DVB-T) as a complementary wireless backhaul/middle-mile connection. .
- *Section IV: Energy-Efficient Networks·* include techniques that describe some ways of minimize the energy conservation, consumption in real time.
- *Section V: Cloud Computing and Resource Management·* describe the issue of resource for energy usage and internet traffic for every device we use in our applications.
- *Section VI: Optimization·* describes the synchronization of the moving devices in the specific regions in order to enable higher reliability in the availability of the requested resources.

## II. COGNITIVE RADIO NETWORKS

CR (cognitive radio) technology paradigm provides communication that exploits efficiently radio spectrum resources. Their technical characteristics based on interactions with the spectrum environment. They can use the un-used frequencies and efficiently access them. This opportunity create new radio spectrum access strategies/policies. Ad-hoc CR networks are characterized by completely self-configuring architectures, where routing is challenging and different from routing in a conventional wireless networks [1]. With this way we don't have delay in transmission based on a signaling mechanism and enabling communication between secondary nodes located in areas with different TV white spaces. Digital technologies have created large markets which data, voice, video and audio are transmitted through wireless and wired networking. [2] Digital television networks have a number of additional frequencies remain unused so as to avoid causing interference between neighboring transmission stations. Secondary CR network operators gain access to unused spectrum, between primary network systems and then operate in a non-interference basis, and locations when and where it is possible for them to transmit, achieving with this way optimum quality of services provided to end users. Thus providing always on triple play services utilize the digital video broadcasting (DVB-T) and avoid interferences or transmission delay [9]. The digital broadcasting is accomplished to digital switch over that after the switch off a big amount of spectrum resources is free for wireless services access / provisioning [8]. Interleaved Spectrum is the large slice of "prime" broadcast frequencies, which remains unused at local level since the analogue TV broadcasting [8]. In many areas with high terrain variations and/or of low population density and dispersed areas national TV broadcasts are either totally absent, or delivered over other medium that VHF/UHF, such as satellites. Effectively, in these areas the unexploited VHF/UHF spectrum is very large, often comprising the entire TV Spectrum. Such geographical White Spaces of TV coverage can be found in rural areas, which tend to be underserved with other broadband options such as Digital Subscriber Line (DSL) or cable. There are techniques that use those frequencies as (DSA) Dynamic Spectrum Access , Spectrum Commons and Spectrum Markers [8].

## III. INTERACTIVE BROADCASTING

- Information and communication technologies (ICT) enable a large number of users worldwide. A lot of companies try have communication with their customers via phone e-mail or internet network. In our days it is a tactic in marketing that internet and worldwide networks have a specially connection between them.

- Interactive marketing allows a company to use direct response to build a relationship with customers [32].

- Marketing is the process used to determine or services customers are interested in and the strategy of what policy should have that the customers be satisfied.

- One of the most development techniques in nowadays is DVB-T. DVB-T is a standard for the transmission of digital TV programs [33]. It is very important cause in our days many applications uses that standard like e-banking and they demand security sensitive information transmission and storage them [33].

- People use more TV set for relax or to forget their routine and get rest. In such case DVB-T have data traffic through several interaction channels [32].

- The DVB-T standard provides the ability to combine digital television and IP data in one UHF channel and broadcast that to users over a wide area. Because of its broadcasting nature, protection of the sensitive data transmitted encapsulated in the DVB-T MPEG transport stream is needed. Without this protection any adversary can receive the data intended for a user by just using a laptop and a DVB-T receiver card with appropriate software [26].

- Furthermore, television radio spectrum, which is a range of low frequencies in VHF and UHF bands, is traditionally used exclusively by analogue television broadcasters [7].

- There is a global move to switch over from analogue to digital television.

- This is called digital switchover (DSO), or in some cases, the analogue switch (ASO), referring to the time when digital terrestrial broadcasting begins, or when analog broadcasting ends, respectively [7]. Some of the spectrum bands used for analogue television will be totally cleared and made available for usage by other wireless networks. Market analysis shows that the "digital dividend" is a unique opportunity to realize economic/social benefits across countries [7].The transmission from analog to digital will free valuable spectrum to increase reliability and efficiency broadcasting.

- There is another one technique called DVB/IP backhaul environment that enables users to access triple-play IP services at a guaranteed QoS. Users receive/deliver triple-play IP-services via intermediate communication nodes, which make use of wired/wireless technologies in the access network. Nowadays, backhaul networks are mainly based on wireless technologies (e.g. Microwave), providing for the fast deployment of broadband metropolitan networking infrastructures within any terrain, besides reducing the high installation costs usually required for a wired/cable backhaul connection [30].

- It proposes a network architecture that exploits the terrestrial digital video broadcasting technology (DVB-T) as a complementary wireless backhaul/middle-mile connection. DVB-T technology is not only used for delivering custom linear content (i.e. one -way services) but also as a medium for the provision of interactive multimedia and on-demand services [30].

# IV. Energy-Efficient Networks

As years passes global industrial applications and devices (personal computers, mobile, smart-phones, tablets) become more efficiency and more usefully created a new field that propose mechanisms, which consider the energy consumption of the wireless devices, while running an application. This make the need for a dependable mobile computing processing environment. These devices while are in use of sharing resources and data loss their network connectivity as they moved. Therefore, a mechanism that faces the connectivity problem and enables the devices to react in frequent changes in the environment, while it enables energy conservation, is of great need. This mechanism will positively facing the wireless loss resources. For this reason we must make estimation and find an estimation model which reset us in previous condition. One of this model-mechanism technique called backward estimation model and facing the availability, based on the traffic difference through time and the associated traffic characteristics difference in a model offering Energy Conservation and minimization of the wireless requested resources [41]. This mechanism follows the introduced Backward Traffic Difference (BTD) scheme [42]. The BTD scheme initially evaluates the data off traffic and according to the delay limits the model adds a second level of traffic difference. Another one mechanism-model is that who takes into consideration the energy consumption of the wireless devices, while running an application. That model is considered as "low offered" throughput model as that offers low Quality of Service (QoS) or Quality of Experience (QoE). This mechanism technique is as a part of the run application start up minimizing the GPU/CPU efforts and the energy of the device is running out of resources.

For the Backward Traffic Difference structure follows Section II describes the related work done and the need in adopting a Traffic-based scheme, and then Section III follows by presenting the proposed.

Backward Traffic Difference estimation for Energy Conservation. Section IV presents the real time performance evaluation results focusing on the behavioral characteristics of the scheme and the Backward Traffic Difference along with the system's response, followed by Section V with the conclusions and foundations, as well as potential future directions.

For the "low offered" throughput model related work and the research motivation are described in Section II. Section III proposed offloading scheme and the associated mechanisms to reduce the energy consumption, maximizing the lifetime of the devices. Section IV presents the results that were obtained, by conducting simulation experiments, towards evaluating the performance of the proposed mechanism, by focusing on the behavioral characteristics of the scheme, along with the system response, as well as the energy consumption achieved. Finally, Section V summarizes the research findings of this paper and discusses the potential future directions for further experimentation and research.

# V. CLOUD COMPUTING AND RECOURCE MANAGEMENT

Nowadays traditional PCs mobile devices (i.e., smart phones, smart tablets, etc.) have become a staple of our society and we have a large number of applications their supported from them but many applications are in need or generate a lot of Internet traffic. Of course, for this to happen, application developers rely on good networking connections with the Cloud. Thus, it is no wonder today that most mobile broadband plans for data access on laptops and tablets or via mobile hotspots are typically tiered. In our work, we started analyzing this fact and found out that today monthly most clients do tend to pay for a lot of broadband traffic, but most of the time they never use. Resource sharing in the wireless and mobile environment is even more demanding as applications require the resource sharing to happen in a seamless and unobtrusive to the user manner, with minimal delays in an unstructured and ad-hoc changing system without affecting the user's Quality of Experience (QoE).

Tiered data based on how much data the client uses per month. Networks applications have to overcome distances and connect people in every location around the world. Especially provide users with their location, shared information can be stored, find solution of high roaming charges for people who travel enough cost and no precise data application services, users may are in use for a little time. Such problems, smart cities of tomorrow will probably rely on the use of a content sharing service, entirely dependent on mobile devices. Bringing social media and content sharing into ad-hoc networks seems the natural choice towards the next frontier in mobile industry.

The next step is to create a type of ad-hoc network which use broadcast as a method of information distribution. The most important is to find a way that data will stay around until lifetime expires. Internet connections sharing has the ability enables users to create their ad-hoc network with other enabled devices. It is like an automatic sharing based network where the user automatically choose their shared connection network. Those applications need energy. As energy stored in batteries more power applications created every day. For this reason we must found a model of energy usage in every environment and especially in wireless because the most usage internet connections in nowadays are wi-fi. The problem of sharing sources that creates traffic and spend energy and money for many people can be solved by re-routing traffic through other devices. With that way every user could be a traffic provider for other users.

# VII .OPTIMIZATION

The first appearance of the wireless networks was in 1970.Since these networks have been developed so fast. [66]

They must have availability in the requested resources . One of them is Vanet and it used in cars devices to establish a network connectivity. It works for mobile peer to peer devices because exploit the mobile usage and different paths.

The unlimited battery power and storage provided for VANET nodes are a key advantage that suffer from energy limitations, thus, the network lifetime is longer and it can store as many data as needed [71].

VANETs are used for safety and comfort of passengers inside the vehicles, increase the liability of the vehicles when accidents occur, and reduce the traffic congestion [71].

This work exploits the movements of the devices and the passive device synchronization to increase end-to-end file sharing efficiency through users and Mobile Infostations [67].

In last decade all applications moved to wireless networks and the mobile wireless network start to grow up .It called and ad hoc network .One ofthem is MANET . A MANET is a set of mobile nodes that communicate through wireless links. Each node is like a host and router. As nodes are mobile devives, the network can change rapidly and unpredictably over time[66].Because of that a MANET does not have base station and communicate between nodes with the wireless links. MANETs is the type of networks used such as military operation, disaster recovery, and temporary conference meetings.[66]. MANET have specific type of algorithm that implement the communication.

The new on the move applications sharing resources of each device in a new technology that uses moving cars as devices/nodes in a dynamically changing network to establish a mobile network connectivity [70].

It takes advantages of the movements of cars-devices and the passive device synchronization to increase end to end file sharing. [70]

Simulation results have shown that scheme offers high throughput and reliability and a solution for sharing resources of any capacity in dynamically changing mobile peer-to-peer wireless network.

Wireless networks are used in many real-time applications that offer specialized services 'on-the-move', where these services require reliable communication and continuous end-to-end connectivity [70]

It takes advantages of the movements of cars-devices and the passive device synchronization to increase end to end file sharing [70].

Simulation results have shown that scheme offers high throughput and reliability and a solution for sharing resources of any capacity in dynamically changing mobile peer-to-peer wireless network.

Wireless networks are used in many real-time applications that offer specialized services 'on-the-move', where these services require reliable communication and continuous end-to-end connectivity [62].Devices have many problems with that demand of always on from users and they unable to handle resource sharing.

A model solve this problem is to combine the resource share characteristics with the opportunistic content sharing procedure in order to offer higher reliability of the requested sources [62].